\documentclass[aps, reprint,showpacs,showkeys,pra,nofootinbib]{revtex4-1}
\usepackage{dcolumn}
\usepackage[utf8]{inputenc}
\usepackage{amsmath}
\usepackage{amsfonts}
\usepackage{amssymb}
\usepackage{graphicx}
\usepackage{hyperref}
\usepackage{xcolor}

\begin{document}

\title{Inconsistency of a realistic interpretation of quantum measurements: A simple example}

\author{Pablo L. Saldanha}\email{saldanha@fisica.ufmg.br}
\affiliation{Departamento de F\'isica, Universidade Federal de Minas Gerais, Caixa Postal 701, 30161-970, Belo Horizonte, MG, Brazil}

\date{\today}

\begin{abstract}
We use a simple example to illustrate why it is not possible to consider that a measurement reveals an underlying objective reality of a property of a quantum system, that continues the same after the measurement is performed. This kind of incompatibility between realism and quantum mechanics is theoretically demonstrated with an example where sequential spin measurements are performed on a spin-2 quantum particle. We discuss the relation of this result with other works that investigate the concept of reality in quantum mechanics.
\end{abstract}



\maketitle

\section{Introduction}

Investigations about the possibility or not of a realistic interpretation of quantum phenomena, in which one could ascribe an objective reality to properties of a quantum system, generated intense and fruitful debates in the past 90 years. The Einstein-Podolsky-Rosen (EPR) paradox \cite{einstein35,bohr35}, Bell inequalities \cite{bell64,bell,mermin85,dietrich02,aspect02}, quantum contextuality \cite{kochen67,mermin90,cabello14,peres}, investigations about the reality of a quantum state \cite{pusey12,lewis12,leifer14,barrett14,ringbauer15,colbeck17} and other discussions \cite{peres,laloe01,despagnat,ballentine,griffiths17,griffiths19} continuously fueled such debate. The recent experimental violation of Bell inequalities simultaneously closing the detection efficiency and locality loopholes \cite{hensen15,giustina15,shalm15} show that, under certain reasonable assumptions such as free will and absence of superdeterminism, there can be no local realistic description of Nature. Quantum contextuality, on the other hand, shows that it is impossible to simultaneously ascribe definite objective values for all quantities that can (in principle) be measured in quantum systems \cite{peres}.

Here we extend this discussion by illustrating why, according to the quantum mechanical laws, it is not possible to consider that a measurement reveals an underlying objective reality of a property of a quantum system prior to the measurement procedure, that continues the same after the measurement is performed. In other words, if the we measure the $z$ component of the spin of a particle $S_z$ and find the value $2\hbar$, it is not possible to consider that we have $S_z=2\hbar$ objectively \textit{in the real world} just before the measurement is performed and just after it. This behavior completely contradicts the classical view of measurements, where their purpose is exactly to obtain information about the properties of a system, that would exist independently of the act of measuring them. This result is demonstrated below in a simple example where sequential spin measurements are performed on a spin-2 quantum particle. We also discuss here the relation of this result with other investigations about the possibility or not of a realistic interpretation of quantum phenomena. 

\section{The Example}

Consider the sequential spin measurements performed on a spin-2 particle represented in Fig. 1. If a spin-2 particle enters the first Stern-Gerlach apparatus and is not detected by the (assumed perfect) screen detector positioned at the exits correspondent to $x$ spin component between $-2\hbar$ and $\hbar$, the only possibility is that it has followed the path corresponding to the value $2\hbar$ for the $x$ spin component. The absence of particle detection on the screen effectively acts as a spin measurement, and the system state after this measurement can be written as 
\begin{equation}\label{state}
	|2\rangle_x=\frac{1}{4}|2\rangle_z+\frac{1}{2}|1\rangle_z+\frac{\sqrt{6}}{4}|0\rangle_z
	                 +\frac{1}{2}|-1\rangle_z+\frac{1}{4}|-2\rangle_z, 
\end{equation}
where $|m\rangle_i$ represent an eigenstate of $S_i$ (the $i$-th component of the particle spin) with eigenvalue $m\hbar$ \cite{cohen}. So there is a nonzero probability that a measurement of $S_z$ performed  on a particle in this state results in $2\hbar$, as depicted in Fig. 1. In this paper we criticize possible realistic interpretations about the system ontological physical state at the time between these two measurements. 

\begin{figure}
  \centering
    \includegraphics[width=8.9cm]{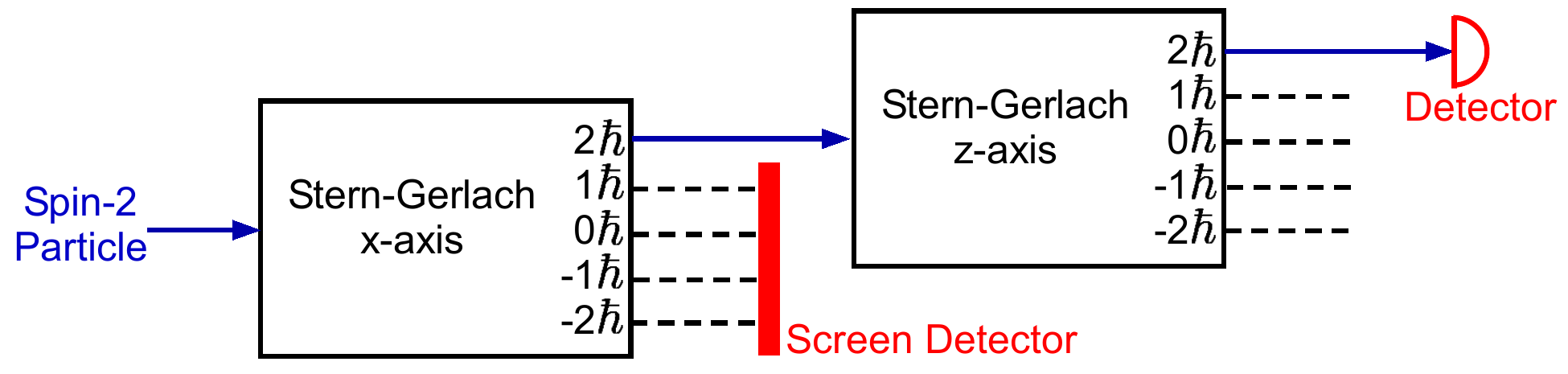}
  \caption{Sequential Stern-Gerlach measurements performed on a spin-2 particle. A measurement of $S_x$ results in $2\hbar$ and a sequential measurement of $S_z$ also results in $2\hbar$.}\label{fig:SG}
\end{figure}

As we show in the following, we have to abandon at least one of the two following assertions related to the system ontological state between the two Stern-Gerlach measurements depicted in Fig. 1: 
\begin{itemize}
	\item (a) after the measurement of $S_x$ we have an objective reality associated to the value obtained in the experiment; or
	\item (b) a measurement of $S_z$ reveals a pre-existing value for this quantity.
\end{itemize}
The reason is simple: If both assertions above are true, we conclude that $S_x^2=S_z^2=4\hbar^2$ at a time between the measurements in the treated example, such that
\begin{equation}\label{sum}
	S^2=S_x^2+S_y^2+S_z^2\ge 8\hbar^2,
\end{equation}
since the possible values that $S_y^2$ can assume are positive or zero. However, for a spin-2 particle a measurement of $S^2$ must result in $6\hbar^2$, what leads us to the conclusion that Eq. (\ref{sum}) is incompatible with quantum mechanics. So at least one of the two assertions (a) or (b) stated above must be false. 

In the example of Fig. 1, the result of the first measurement  (of $S_x$) is naturally consistent with assertion (a), since sequential measurements of $S_x$ would certainly result in this same experimental value. However, the second measurement (of $S_z$) is naturally consistent with assertion (b), since we associate a particular particle deflection with a particular value for the spin component along the magnetic field direction because we can calculate, using the electrodynamics laws, what is the spin component that gives rise to this particular particle deflection in the Stern-Gerlach apparatus. On this way, it is very interesting to note that both assertions cannot be right, such that we must abandon our classical view of the measurement procedure in quantum systems. Also, it is clear that we cannot assume that the particle has a magnetic moment pointing in a particular direction, but that the quantum mechanical formalism is unable to tell which direction, since this view would lead to the validity of Eq. (\ref{sum}) in the treated example, contradicting the quantum mechanical predictions.

Similar paradoxes arise with particles with other spin quantum numbers $s$. If two sequential measurements of orthogonal components of the particle spin are found to have values with the maximum possible modulus $s\hbar$, the paradox appears if the sum of these values squared is greater than the value of $S^2$, i.e., if $2s^2\hbar^2>s(s+1)\hbar^2$. So we may have a paradox whenever $s>1$. $s=3/2$ is the lower dimension case, but we've chosen an example with $s=2$ for aesthetic reasons, avoiding the presence of too many fractions along the text.

\section{Discussion}

The word `realism' may have different meanings when applied to physical theories, and in the following we discuss how our results affect the possibility or not of a realistic interpretation of quantum mechanics in different contexts for the expression `realistic'. 

It is usual to say that, in view of the Bell inequalities, the EPR notion of local realism is incompatible with quantum mechanics. But by considering the presented results we can say that that, with the assumption that measurements reveal an independent ontological value for a property of a quantum system, the EPR criterion of physical reality (with no mention to locality)  would be already incompatible with quantum mechanics. This criterion states that ``if, without in any way disturbing a system, we can predict with certainty (i.e., with probability equal to unity), the value of a physical quantity, then there exists an element of physical reality corresponding to this physical quantity'' \cite{einstein35}. This criterion leads directly to assertion (a), since the state of Eq. (\ref{state}) is in an eigenstate of $S_x$, such that the value of the physical quantity associated to this observable can be predicted with certainty, without further disturbance. So we conclude that the EPR criterion of physical reality is inconsistent with assertion (b), such that at least one of them must be abandoned.

In the context of Bell inequalities \cite{bell64,bell,aspect02,hensen15,giustina15,shalm15}, the expression `realism' is associated to the existence of hidden variables $\lambda$, not considered in the quantum theory, that would determine (possibly in a probabilistic manner) the values that would be obtained in the measurements of physical properties of the system. A similar `hidden variables' view is adopted on the recent discussions about if the quantum state corresponds directly to reality or if it represents our knowledge about a system underlying reality \cite{pusey12,lewis12,leifer14,barrett14,ringbauer15,colbeck17}. In all these recent works about the reality of the quantum state there is the assumption that ``a system has a `real physical state' not necessarily completely described by quantum theory, but objective and independent of the observer'' \cite{pusey12}. For being consistent with the results presented here, the hidden variables $\lambda$ should determine values for the quantities $S_x^2$, $S_z^2$ and $S_x^2+S_z^2$ independently, with the third variable not being necessarily equal to the sum of the first two. On this way, the quantum description of the spin angular momentum would be only an apparent manifestation of the hidden variables, not a fundamental description of the system. On these terms, we can have a realistic description of the system independently of the act of measurement, as assumed in Refs. \cite{pusey12,lewis12,leifer14,barrett14,ringbauer15,colbeck17}, by denying an objective reality for the spin angular momentum variables and keeping an objective reality for the hidden variables. But this results in an odd situation. As it was stated before, we can associate different particle deflections with different values for the spin component along the magnetic field direction in a Stern-Gerlach measurement because we can calculate, using the electrodynamics laws,  which particular spin component gives rise to each particular particle deflection. So the unknown hidden variables must act on the particle motion according to unknown physical laws resulting in exactly the same motion that would be obtained with the assumption of the physical existence of the spin component, but without producing a spin component with an objective physical reality. Another option, which we prefer, is to deny the existence of a real physical state, independent of the observer, for quantum systems.

The paradoxical behavior exposed with the treated example is related to von Neuman's argument that for two quantities associated to operators $A$ and $B$ that do not commute, the quantity associated to the operator $A+B$ is not the sum of the quantities associated to $A$ and $B$ separately \cite{neumann,dieks18}. Since $[S_x^2,S_z^2]\neq 0$, the violation of Eq. (\ref{sum}) can be seen as an explicit demonstration of von Neumann's argument. 

The arguments used here are different from the ones used in the  `consistent histories' interpretation of quantum mechanics \cite{griffiths17,griffiths19}. But they lead us to a common conclusion that the classical notion of a measurement, where it reveals the underlying objective reality of the system property being measured, is incompatible with quantum systems.

\section{Conclusion}

In summary, we have demonstrated with a simple example that we cannot consider that a measurement performed on a quantum system reveals the underlying ontological value of the measured quantity, that continues the same after the measurement is performed. To finish, we can provide a response to the famous question formulated by Einstein when discussing the relation between realism and quantum mechanics: ``Do you really believe the moon exists only when you look at it?'' \cite{mermin85}. According to the present results, if the moon is a metaphor for a spin angular momentum component, we should answer that we cannot believe on its independent existence even when we are seeing it shining on the sky. Following Bohr, who believed that quantum mechanics describes quantum systems interacting with measurement apparatuses, not quantum systems themselves, we could say that the moonlight does not show the moon independent existence because the moonlight is not a property of the moon alone, but a property of the moon interacting with our eyes. 

\begin{acknowledgements}

The author acknowledges B\'arbara Amaral, Marcelo Terra Cunha, Rafael Rabelo, Vladimir Hnizdo, and Gerold Gr\"undler for very useful discussions. This work was supported by the Brazilian agencies CNPq, CAPES, FAPEMIG, and CNPq/INCT-IQ (465469/2014-0).

\end{acknowledgements}


%
%



\end{document}